\pdfoutput=1
\documentclass[amsmath,amssymb,pre,twocolumn,floatfix]{revtex4}

\def\D{{\rm d}}

\usepackage{color}
\usepackage{graphicx}
\usepackage{dcolumn}
\usepackage{times}
\usepackage{soul}
\usepackage{cleveref}

\begin{document}

\title{Phase separation and folding in swelled nematoelastic films}
\author{A.~P. Zakharov, L.~M. Pismen}
\affiliation{
Technion -- Israel Institute of Technology, Haifa 32000, Israel}

\begin{abstract}
We explore reshaping of nematoelastic films upon imbibing an isotropic solvent under conditions when isotropic and nematic phases coexist. The structure of the interphase boundary is computed taken into account the optimal nematic orientation governed by interaction of gradients of the nematic order parameter and solvent concentration. This structure determines the effective line tension of the boundary. We further compute equilibrium shapes of deformed thin sheets and cylindrical and spherical shells with the rectilinear or circular shape of the boundary between nematic and isotropic domains.  A differential expansion or contraction near this boundary generates a folding pattern spreading out into the bulk of both phases. The hierarchical ordering of this pattern is most pronounced on a cylindrical shell.
\end{abstract}
 \maketitle

\section{Introduction}

Any change in nematic order in a nematic elastomer film causes it to deform. Following the early theoretical prediction of deformation of monodomain liquid crystal elastomers \cite{degennes75}, its first experimental realization  \cite{Finkelmann81}, and the development of theory combining nematic and elastic contributions to the Landau -- de Gennes functional \cite{Warner88,Khokhlov,Pleiner,Warner}, a variety of shapes in patterned nematoelastic films have been constructed in recent years \cite{Warner12,sharon,Cirak14,epj15,mostajeran2015curvature,mostajeran2016encoding}. The various shapes have been produced by imposing a desired orientation in a liquid-crystalline film prior to polymerization and heating the textured nematoelastic film above the NIT point \cite{Broer14,BroerLa,White,WhiteSM}. Textures can be also affected by composition changes carried out either by adding a suitable dopant (most commonly, by light-induced isomerization \cite{Finkelmann91,Ikeda,Samitsu,Broer}) or by swelling the nematic gel in either nematic \cite{UrayamaN} or isotropic solvents \cite{Urayama03,Urayama06}.  

Swelling of either isotropic or nematic gels is counteracted by entropic elasticity of the polymer network \cite{Bahar}, and equilibrium swelling is governed by the balance of the mixing and elastic energy.  An additional factor in nematic gels is the change of nematic order, which is reduced when an isotropic solvent is imbibed. Recently, Cheewaruangroj and Terentjev \cite{Terentjev2015} brought attention to an interesting possibility of \emph{coexistence} of a monodomain nematic state at lower, and the isotropic one at higher solvent concentrations. These authors investigated the coexistence conditions for nematic elastomers in one-dimensional (1D) string geometry, as they write, ``especially to avoid complicate issues of inhomogeneous swelling". It is exactly these more complicated phenomena arising in two-dimensional (2D) film geometry that we wish to explore in this communication. 

After formulating the basic computation procedure in Sect.~\ref{S1}, we explore in Sect.~\ref{S21} the properties of the boundary (front) between the nematic and isotropic state, assuming the overall energies of both states, comprising elastic, nematic, and mixing constituents, to be equal, and the front to be stationary and rectilinear. Next, we explore, with the help of the algorithm delineated in Sects.~\ref{S13}, \ref{S22}, deformation of a nematic film or shell  brought into contact with a solvent. We do not consider a very complicated dynamic process of inhomogeneous swelling, phase separation, and coarsening of an emerging pattern but concentrate upon the final equilibrium state minimizing the overall energy of the system, including the energies of separated nematic and isotropic domains and the energy of their boundary. Since the width of the latter, largely determined by nematic elasticity, is small, it can be treated in macroscopic computations as a line characterized by the line energy extracted from the preliminary computation.  We shall see that differential expansion or contraction near the front generates a folding pattern spreading out into the bulk of both phases. The folding patterns are substantially different in flat sheets (Sect.~\ref{S31}) and cylindrical (Sect.~\ref{S32}) or spherical (Sect.~\ref{S33}) shells. 

\section{Basic equations \label{S1}}
\subsection{Nematic alignment}\label{S11}

The nematic alignment in a flat film is obtained by minimizing the nematic energy functional $\mathcal{F}_{n}$ dependent on the 2D tensor nematic order parameter \textbf{Q} expressed in Cartesian coordinates as
\begin{equation}
\mathbf{Q} = \left(\begin{array}{cc}p & q \\ q & -p \end{array} \right)
= \frac {S}{\sqrt{2 }}\,\left(\begin{array}{cc}\cos 2\vartheta & \sin 2\vartheta \\ \sin 2\vartheta & -\cos 2\vartheta \end{array} \right),
\end{equation} 
where $S=\sqrt{\mathrm{Tr}(\mathbf{Q}\cdot\mathbf{Q})}=\sqrt{p^2+q^2}$ is the scalar nematic order parameter (NOP), and $\vartheta$ is the director orientation angle. 
The nematic energy per unit area in a layer with the solvent fraction $\varphi$ is expressed as ${\cal F}_\mathrm{n}=\int {\cal L}_\mathrm{n}\D^2 \mathbf{x}$ with the Lagrangian
\begin{align}
&{\cal L}_\mathrm{n} =  (1-\varphi) \alpha h\biggl[-\frac{1-\gamma \varphi}{2}Q_{ij}Q_{ij}  +
\frac{1}{4}\left(Q_{ij}Q_{ij} \right)^2  \biggr.\notag \\
 & \biggl. + \frac{\kappa_1}{2} \left|\nabla_i Q_{ij}\right|^2 
  + \frac{\kappa_2}{4} \sum_{ijk}\left(\nabla_iQ_{jk}\right)^2
 -\beta \nabla_i \varphi \nabla_j Q_{ij} \biggr],
 \label{Ln}
\end{align}
where $\alpha$ is the characteristic nematic orientation energy per unit volume, $\gamma$ is a parameter quantisizing the dependence of nematic order on the solvent fraction, $h$ is the film thickness, and $\kappa_1, \kappa_2$ are elastic constants (not taking into account their dependence on nematic orientation); summation over repeated indices is presumed. This expression contains the dependence on the solvent ratio $\varphi$ and its gradient. The gradient terms are important only in the vicinity of defects and phase boundaries but the term mixing the gradients of the NOP and solvent ratio \cite{Fukuda,rey04,kopf2013phase,arxivorg} plays an important role by setting nematic orientation near the interphase boundary. The effect of this term is made transparent when only the scalar NOP $S$ is variable. With fixed orientation, Eq.~\eqref{Ln} simplifies to
\begin{align}
{\cal L}_\mathrm{n} = &  (1-\varphi) \alpha h\biggl[-\frac{1-\gamma \varphi}{2}S^2 +
\frac{1}{4}S^4  + \frac{\kappa}{2} (S_x^2 +S_y^2 ) \biggr.\notag \\
 - & \biggl. \beta (\varphi_xS_x \cos 2\vartheta +\varphi_xS_y\sin 2\vartheta)
\biggr],
 \label{LnS}
\end{align}
where $\kappa=\kappa_1+\kappa_2$ and indices denote partial derivatives, with $\varphi$ assumed to be changing along the $x$ axis. Assuming that $S$ decreases with $\varphi$ ($\gamma>0$), the optimal angle, reducing the overall anergy at $\beta<0$,  is $\vartheta=0$, so that the director tends to orient along the concentration gradient. At $\beta>0$, on the opposite, the lowest energy is attained at $\vartheta=\pi/2$, when the director is oriented normally to the gradient (along the $y$-axis); the change of  $S$ along the $y$-axis does not affect this argument. This term is therefore responsible for \emph{spontaneous anchoring} at a nematic-isotropic interface \cite{kopf2013phase,arxivorg}.

\subsection{Swelling and nematic-isotropic demixing} \label{S12}

Imbibition of a solvent changes the energy of an isotropic gel, on the one hand, by the entropic effect of mixing and, possibly, gel-solvent van der Waals or polar interactions, and, on the other hand, by stretching the polymer network. The free energy of solvent-polymer interaction $\mathcal{F}_\mathrm{m}=\int\mathcal{L}_\mathrm{m}\D^2 \mathbf{x}$ is given by the Flory-Huggins equation \cite{Flory}
\begin{align}
{\cal L}_\mathbf{m}  &=  h \biggl[\frac{\chi}{2} \left|\nabla \varphi \right|^2 - \zeta  \varphi (1-\varphi)
 \notag \\
  &  +   nkT [ \varphi \ln\varphi + (1-\varphi)\ln(1-\varphi) ]  \biggr],
 \label{Fmix}
\end{align}
where $\chi>0$ is the rigidity coefficient,  $ n$ is the total number of solvent molecules and monomer segments per unit volume,  $k$ is the Boltzmann constant, $T$ is temperature, and $\zeta$ is the Flory-Huggins interaction parameter. 

The \emph{internal} elastic energy of a swelling isotropic gel is \cite{Terentjev2015} ${\cal F}_e  =  \frac{3}{2} n_fkT V_0 (1-\varphi)^{-2/3}$, where $n_f$ is the number of network filament segments per unit original volume $V_0$. In a nematic gel, swelling causes, in addition, a change of nematic energy, and the equilibrium state should be determined by minimizing the sum of nematic, mixing, and elastic energies, with the latter acquiring a much more complicated form dependent on the change of NOP.  

In a uniform (either monodomain nematic or isotropic) state,  the scalar NOP is rigidly tied to the solvent fraction, so that, according to Eq,~\eqref{LnS}, $S= \sqrt{1-\gamma \varphi}$ or $S=0$ at $\varphi>\varphi_c= 1/\gamma$. If the nematic orientation remains fixed in the course of solvent imbibition, and only the scalar order parameter $S$ decreases, the length shortens upon nematic-isotropic transition along the director by the factor $\lambda=1+ a(1-\sqrt{1-\gamma\varphi})$ and extends in the two normal direction by the factor $\sqrt{\lambda}$. As a result, the elastic swelling energy becomes anisotropic, and the energy of a uniform film reduces to 
\begin{equation}
{\cal F}_\mathrm{e}  =  
 \frac{n_fkT V_0}{2(1-\varphi)^{2/3}}\left(\frac{1}{\lambda^2}+2\lambda \right).
   \label{eq:cstel} \end{equation}
The optimal swelling of a uniform film is obtained by minimizing ${\cal F}_\mathbf{e} +{\cal F}_\mathbf{m}$ with respect to $\varphi$ with the gradient term in Eqs.~\eqref{Fmix} omitted and $S$ and, consequently, $\lambda$ related to $\varphi$ as stated above.
Nematic-isotropic demixing takes place when there are two energy minima \cite{Terentjev2015}, one corresponding to the isotropic state at $\varphi \geqslant1/\gamma$ and the other, to the nematic state at $\varphi < 1/\gamma$. 

\subsection{Deformation and bending} \label{S13}

Nematic--isotropic transition (NIT), as well as any change of NOP, necessarily causes the initially flat film to buckle, since in-shell deformations are strongly discouraged in thin films, and it is otherwise impossible to accommodate the change of metric caused by extension and shortening along and across the director. Bending, in turn, may affect the distribution of the solvent fraction. 

The \emph{macroscopic} elastic energy of a film is determined by deviations $\mathbf{u}, \,v$ from, respectively, optimal local in-shell and thickness deformations, and  is defined as $\mathcal{F}_\mathrm{e}= \int {\cal L}_\mathrm{e}\D^2 \mathbf{x} $ with 
\begin{equation}
\mathcal{L}_\mathrm{e} = \frac {E}{2} \left[ |\mathbf{u}|^2
+v^2 +  \frac { h^2}{9}  \mathrm{Tr}( \mathbf{C}^2)  \right] .
 \label{Le}
\end{equation}
where $E$ is the Young modulus and $\mathbf{C}$ is the curvature tensor; the coefficient at the curvature term corresponds to the Poisson ratio 1/2. This functional is discretized on a domain triangulated by the Delaunay algorithm \cite{Delaunay} as
\begin{align}
\mathcal{F}_{e} &= \frac {n_fkT }{2} \sum_\mathrm{nodes} \Biggl[ V_i \Biggl(\frac{1}{2} \sum_\mathrm{adj.n.}\left(\frac{l_{ij} }{\overline{l}_{ij}}-1\right)^2  \notag \\
  & + \left(\frac{h_i }{\overline{h}_{i}}-1\right)^2
+ \frac{h_i^2 C_{i}^2}{9} \Biggr) \Biggr].
\label{Fg}
\end{align}
Here $V_i=h_i A_{i}$ is the instantaneous volume at a node $i$, where $A_{i}=\langle A_{ij}\rangle$ is the average surface area of the adjacent tiles. The local thickness values $h_i$, as well those of $\varphi$ and $\lambda$, are defined at nodes, and the local curvature at a node is computed as $C_i^2= 4H_i^2-2K_i$, where the Gaussian curvature $K_i=(2\pi-\sum \rho_j)/A_i$ is expressed through the angles between two adjacent links $\rho_j=\angle(l_{ij},l_{ij+1})$, and the mean curvature $H_i=\sum (l_{ij}\eta_j)/(4A_i)$, through the angles between the normals to neighbouring triangles, $\eta_j=\angle(\mathbf{m}_{ij},\mathbf{m}_{ij+1})$. The first term accounts for deviations of the observed length $l_{ij}$ from the "optimal" length $\overline{l}_{ij}$ accounting for the intrinsic elongation or shortening following swelling and NIT. The term is divided by 2 because each edge $l_{ij}$ is counted twice in the sum over all nodes. The in-shell length transformation matrix due to NIT for an edge at an angle $\psi$ to the director is $\mathbf{R}^{-1}(\psi)\boldsymbol{\Lambda}\mathbf{R}(\psi)$, where $\boldsymbol{\Lambda}$ is the diagonal matrix with the elements $\{1/\lambda, \sqrt{\lambda} \}$ and $\mathbf{R}(\psi)$ is the rotation matrix. After adding the swelling effect and accounting for the change of the thickness $h$, this yields
\begin{equation}
\begin{gathered}
\overline{l} =\dfrac{l_0 \sqrt{\frac 14\left(\sqrt{\lambda }-\frac{1}{\lambda}\right)^2  \sin^2 2\psi + 
		\left(\sqrt{\lambda }\sin ^2 \psi+\frac{\cos^2 \psi}{\lambda} \right) ^2}}{(1-\varphi)^{1/3}}, \\
\overline{h} =\dfrac{h_0 \sqrt{\lambda}}{(1-\varphi)^{1/3}}.
\label{length}
\end{gathered}
\end{equation}
The expressions \eqref{Le}, \eqref{Fg} do not contain a bulk modulus, since the gel can be viewed as incompressible at a constant solvent ratio. The volume can, however, change due to expulsion or imbibition of the solvent, which requires recomputing $\varphi$ by minimizing Eqs.~\eqref{Fmix}, \eqref{eq:cstel} discretized on the same mesh in conjunction with Eq.~\eqref{Fg} (see Sect.~\ref{S32} for more detail).

\section{Numerical computation}

\subsection{Interphase boundary}\label{S21}

\begin{figure}[t]
\centering
\begin{tabular}{c}
(a)   \\
\includegraphics[width=0.45\textwidth]{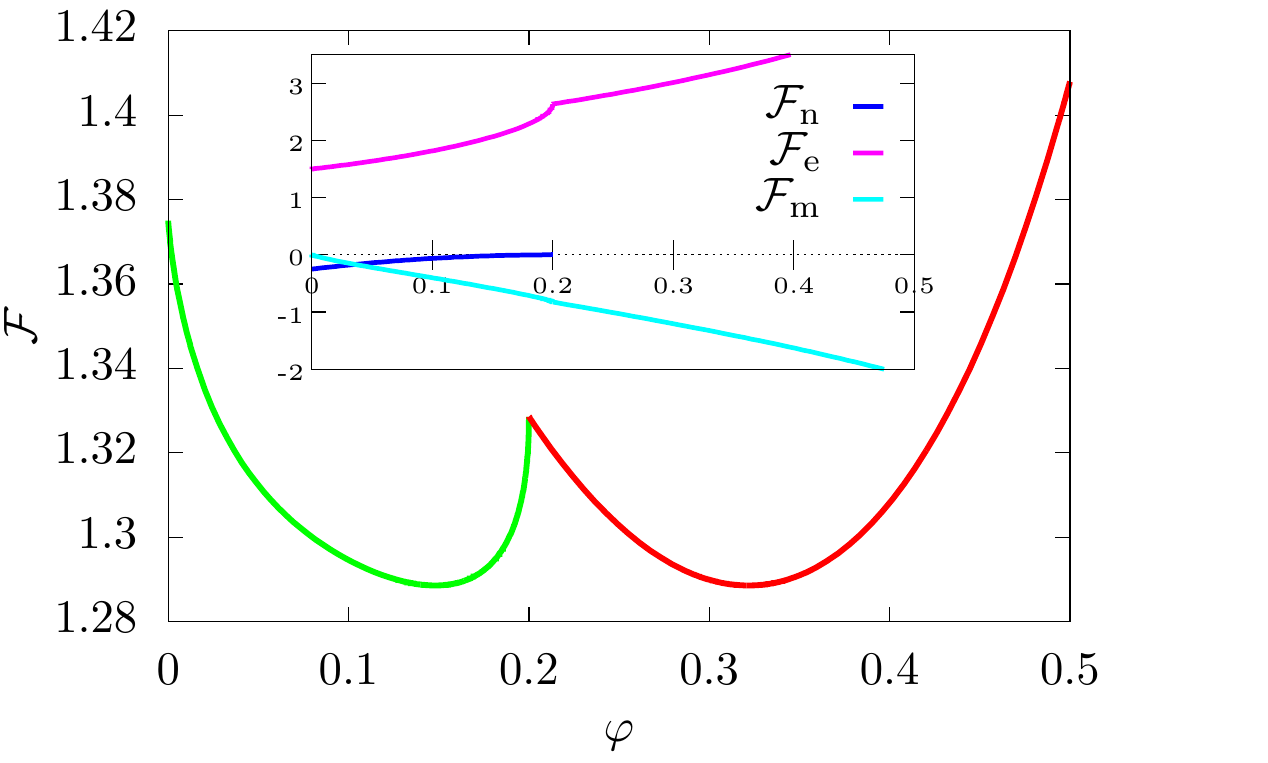} \\
 (b)  \\
\includegraphics[width=0.45\textwidth]{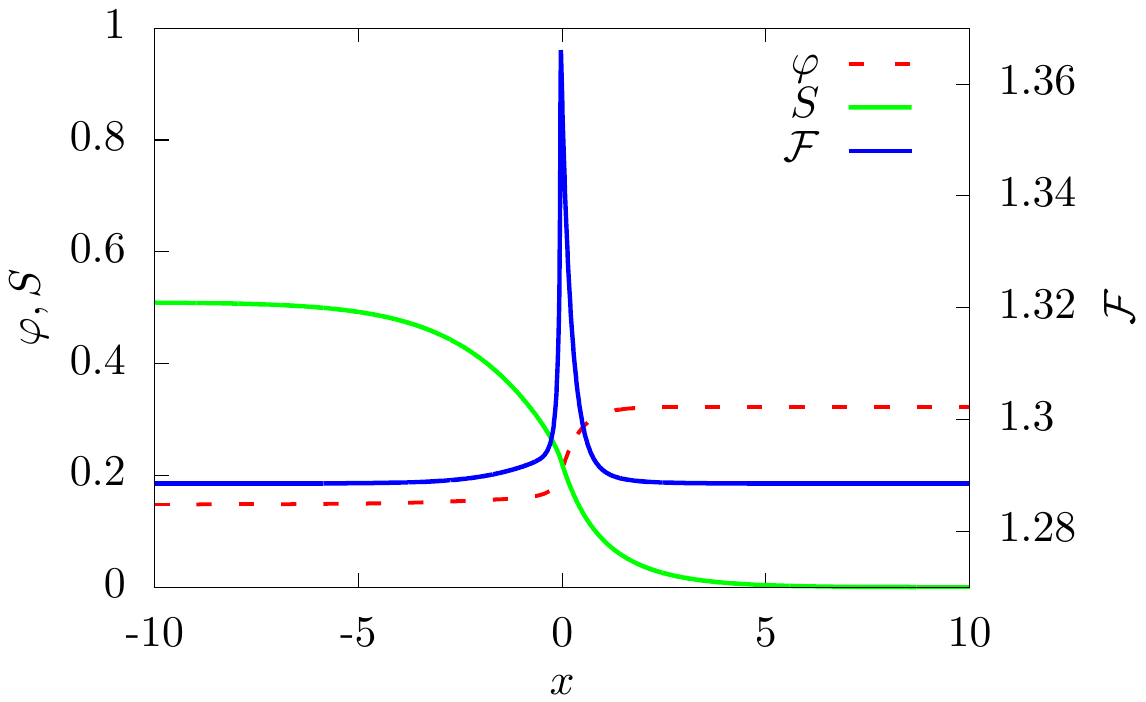}
\end{tabular}
\caption{(a) Dependence of total energy ${\cal F} = {\cal F}_\mathbf{n}+{\cal F}_\mathbf{e}+{\cal F}_\mathbf{m}$ and respective nematic, elastic, mixing energies (in the inset) on the solvent fraction $\varphi$. (b) The dependence of the scalar NOP $S$ and $\varphi$ on the coordinate $x$ normal to the front.}
\label{fig:2minima}
\end{figure}

 We are interested in the case when the isotropic and nematic states coexist, being separated by an interphase boundary, or a \emph{front}. In an undeformed film, the front is expected to be stationary and rectilinear when the sum of mixing and elastic energies,  Eqs.~\eqref{Fmix}, \eqref{eq:cstel} in the uniform nematic and isotropic states have two equal minima.  This happens at certain combinations of parameters, and was fixed in our computations by choosing $\alpha=0.5, ~\beta=-0.5, ~\gamma=5, ~n_fkT=1, ~\kappa=1, ~\chi=1, ~\zeta=0$. The dependence of the overall energy and its constituent parts on $\varphi$ for this set of parameters is shown in Fig.~\ref{fig:2minima}a. Under the chosen conditions, the nematic energy plays a minor role, and the overall balance is determined by the mixing and elastic energies, the former being negative and the latter positive, and both increasing their absolute values with $\varphi$, as seen in the inset of Fig.~\ref{fig:2minima}a.

At the front, the gradient terms in Eqs.~\eqref{LnS}, \eqref{Fmix} become important, and the width of the transitional zone is determined by the values of the parameters $\kappa, \, \chi$, and $\beta$. The necessary computation for a flat film or a cylinder with a uniform nematic orientation is one-dimensional (1D). The change of $S$ and $\varphi$ across the front, shown in Fig.~\ref{fig:2minima}b, is independent of the sign of  $\beta$, provided the nematic director is oriented in the optimal way, \emph{i.e.} normally to the front at $\beta<0$, and parallel to the front at $\beta>0 $.  At the ends of the computation interval (far exceeding the front width), $S$ and $\varphi$ approach the values corresponding to the total energy minima in the homogeneous nematic and isotropic phases shown in Fig.~\ref{fig:2minima}a. The interfacial energy (\emph{i.e.} excess over the total energy of the homogeneous state), also shown in Fig.~\ref{fig:2minima}b, exhibits a kink at the front location. The integral of this excess energy determines the line tension of the front responsible for its relaxation to the minimal length (the straight line on a flat sheet or a geodesic on a bent surface).  

\subsection{Deformation due to swelling}\label{S22}

We compute the equilibrium shape of a film, originally in the monodomain nematic state, as it is deformed due to swelling and phase separation into nematic and isotropic domains. The computation follows an iterative minimization procedure on a triangulated mesh, which is refined near the interphase boundary to resolve the local structure. We start with the 1D distribution of $\varphi$ and $\lambda$ obtained above, with the rectilinear interphase boundary oriented in the optimal way, \emph{i.e.} normally or parallel to the director, respectively, at negative or positive $\beta$.  The positions of nodes are then relaxed to minimize $\mathcal{F}_{e}$, following the pseudo-time evolution equations $\partial \mathbf{x}_i/\partial t=-\delta \mathcal{F}_{e} / \delta \mathbf{x}_i$ for the positions of nodes $\mathbf{x}_i$. 

Deformation causes local volume changes,  thereby influencing the solvent fraction $\varphi$, since the gel would squeeze or imbibe the solvent due to in-plane compression or stretching. We update therefore the distribution of $\varphi$ by evolving and $\partial \mathbf{\varphi}_i/\partial t=-\delta (\mathcal{F}_m+\mathcal{F}_e )/ \delta \mathbf{\varphi}_i$ for the local solvent ratios $\varphi_i$ at the nodes. Since deformation and folding takes place at distances large compared to the front width, where the gradients of $\varphi$ and $S$ are small, we neglect the gradient terms in this computation and assume $S$ and $\lambda$ to be rigidly tied to $\varphi$ as in homogeneous domains. We carry out this computation iteratively, relaxing the node positions and solvent fraction in turn until the overall energy minimum is reached. 

Buckling affects the location of the interphase boundary and changes the distribution of the solvent fraction and NOP in its vicinity, so that the front ceases to be rectilinear. The gradients along the front remain, however, small compared to the transverse gradients. The procedure is only slightly modified for non-planar shells, with an appropriately placed front. 

\section{Folding patterns} \label{S3}

 \subsection{A deformed sheet}\label{S31}
 
\begin{figure}[b]
\centering
\begin{tabular}{c}
(a)  \\
\includegraphics[width=0.49\textwidth]{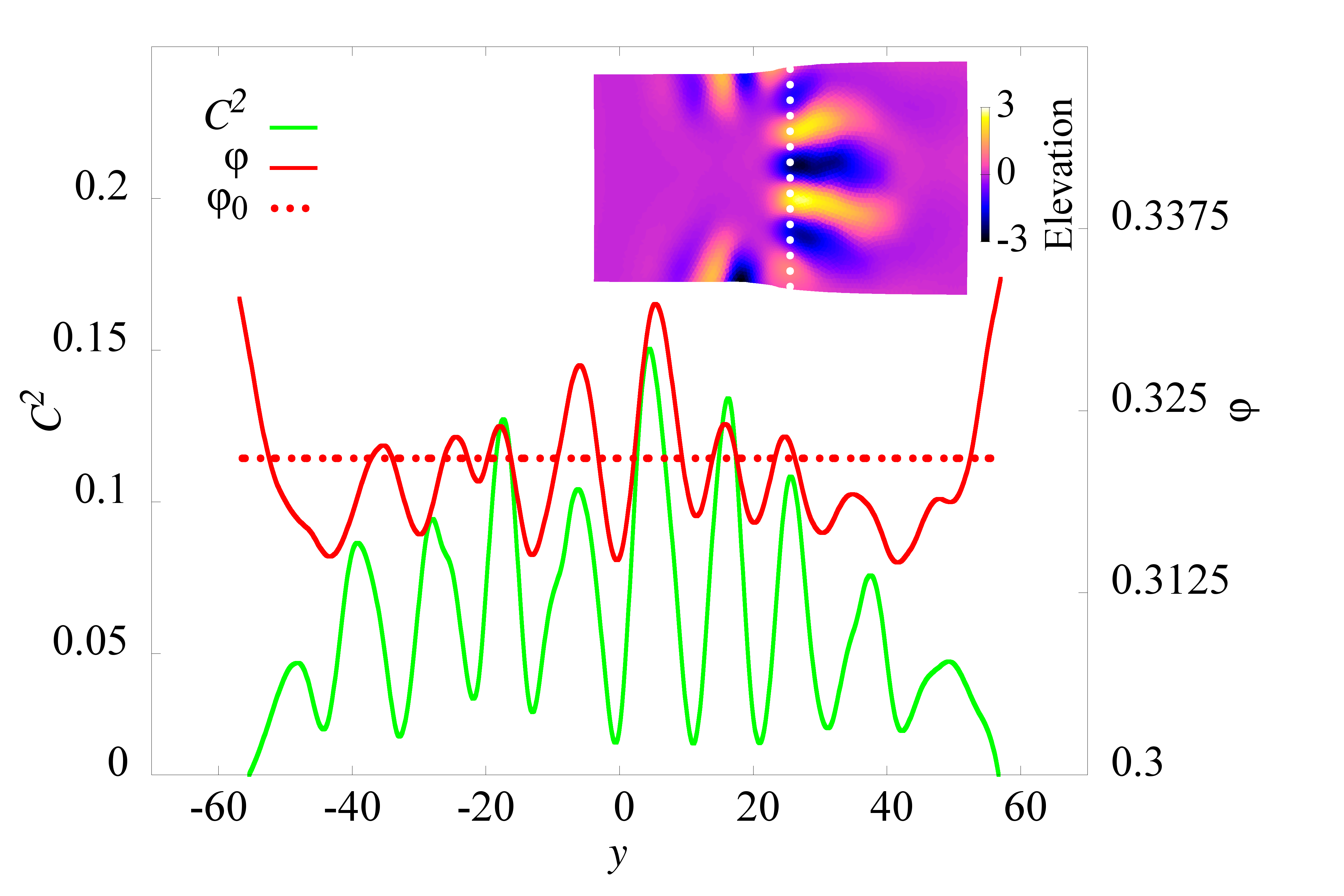}\\
 (b)  \\
\includegraphics[width=0.50\textwidth]{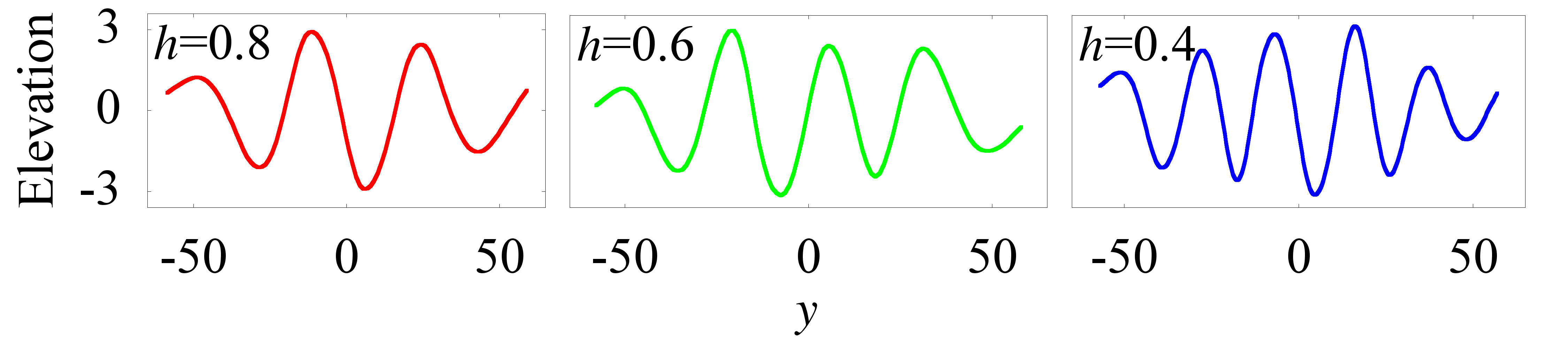}
\end{tabular}
\caption{(a) Squared curvature and solvent fraction along the section parallel to the front at $x=10$ shown by dotted white line in the inset. Inset: the actual shape of a deformed sheet with $h_0=0.8$. (b) Comparison of the profiles along the same section for sheets of different thickness.}
\label{fig:Plane1}
\end{figure}

Folds on a deformed sheet are formed mostly in the isotropic region (on the right side of the inset in Fig.~\ref{fig:Plane1}a), where the anisotropic stretching/compression is more pronounced, while shorter transverse folds are observed near edges in the nematic domain.
 The distribution of the solvent fraction $\varphi$ and vertical deviation from the original planar film are highly correlated, with $\varphi$ larger at highly curved locations. This is seen in the plots of $\varphi$ and squared curvature along a line parallel to the original location of the front (Fig.~\ref{fig:Plane1}a). The number of folds increases with decreasing thickness of the sheet, as seen  in Fig.~\ref{fig:Plane1}b. The folds spread out at distances from the front far exceeding its thickness, and gradually fade away as seen in the inset of Fig.~\ref{fig:Plane1}a.

The folding pattern in Fig.~\ref{fig:Plane1} corresponds to the case $\beta<0$. 
The buckling effect is much weaker at $\beta>0$ when the director is oriented  parallel to the front. Buckling is caused by extension or contraction \emph{along} the front and is initiated due to a large contrast of $S$ and, consequently, the extension ratio $\lambda$ across the front. When the director is oriented normally to the front, both NIT and swelling cause extension parallel to the front, thereby enhancing this effect.  On the opposite, when the director orientation is parallel, contraction due to NIT and swelling partially compensate each other, thereby reducing the effect. The situation is mirrored in the opposite case of isotropic-nematic transition whereby the process is started from the isotropic state and nematic order is established upon drying, with the orientation at the front dependent on the sign of  $\beta$. Then elongation along the front due to parallel orientation is compensated by shrinking, while shortening along the front due to normal orientation is, on the opposite, enhanced. 
 
 \subsection{Folds on a cylinder}\label{S32}

\begin{figure}[b]
	\begin{tabular}{cc}
	\includegraphics[width=0.22\textwidth]{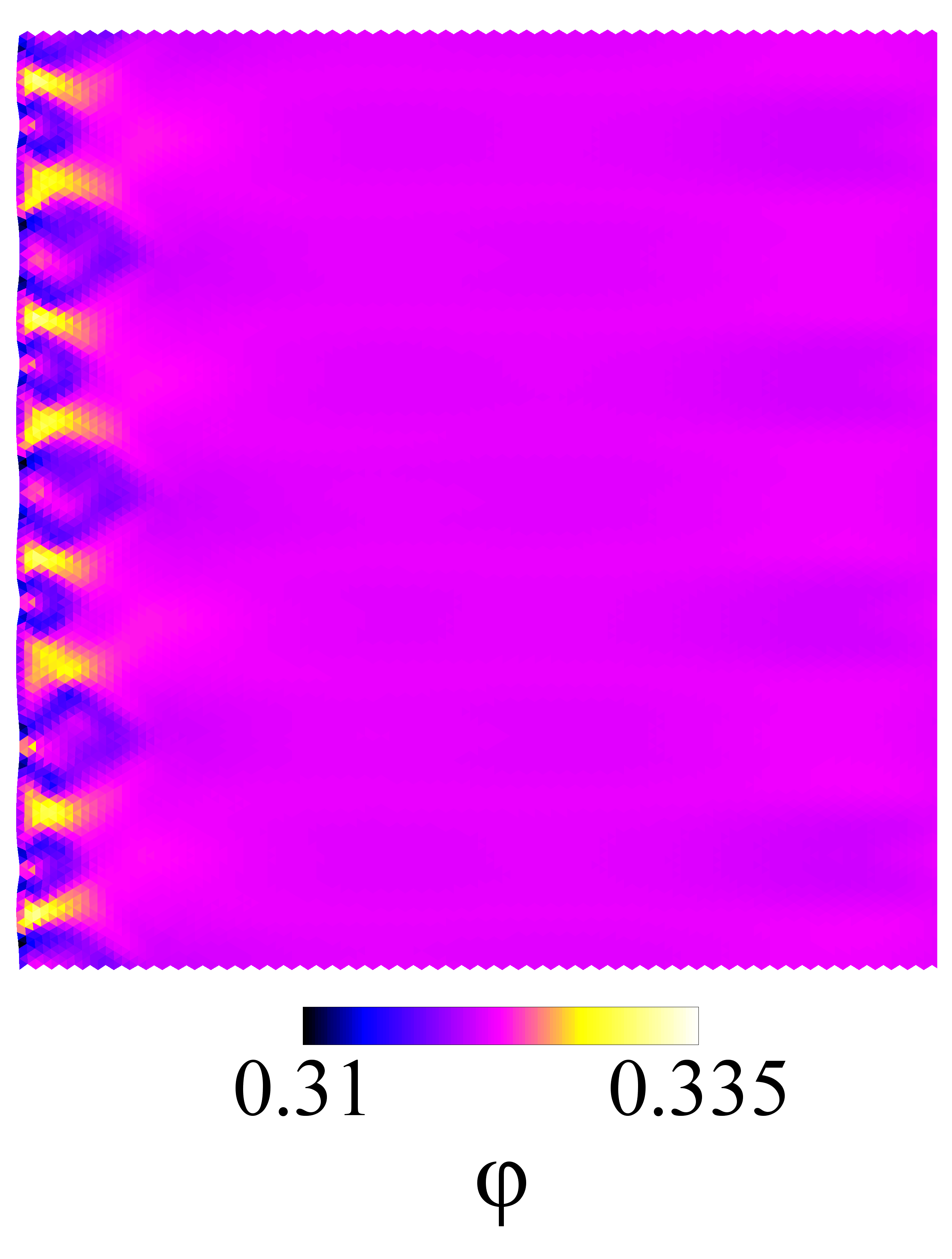} 
	&\includegraphics[width=0.22\textwidth]{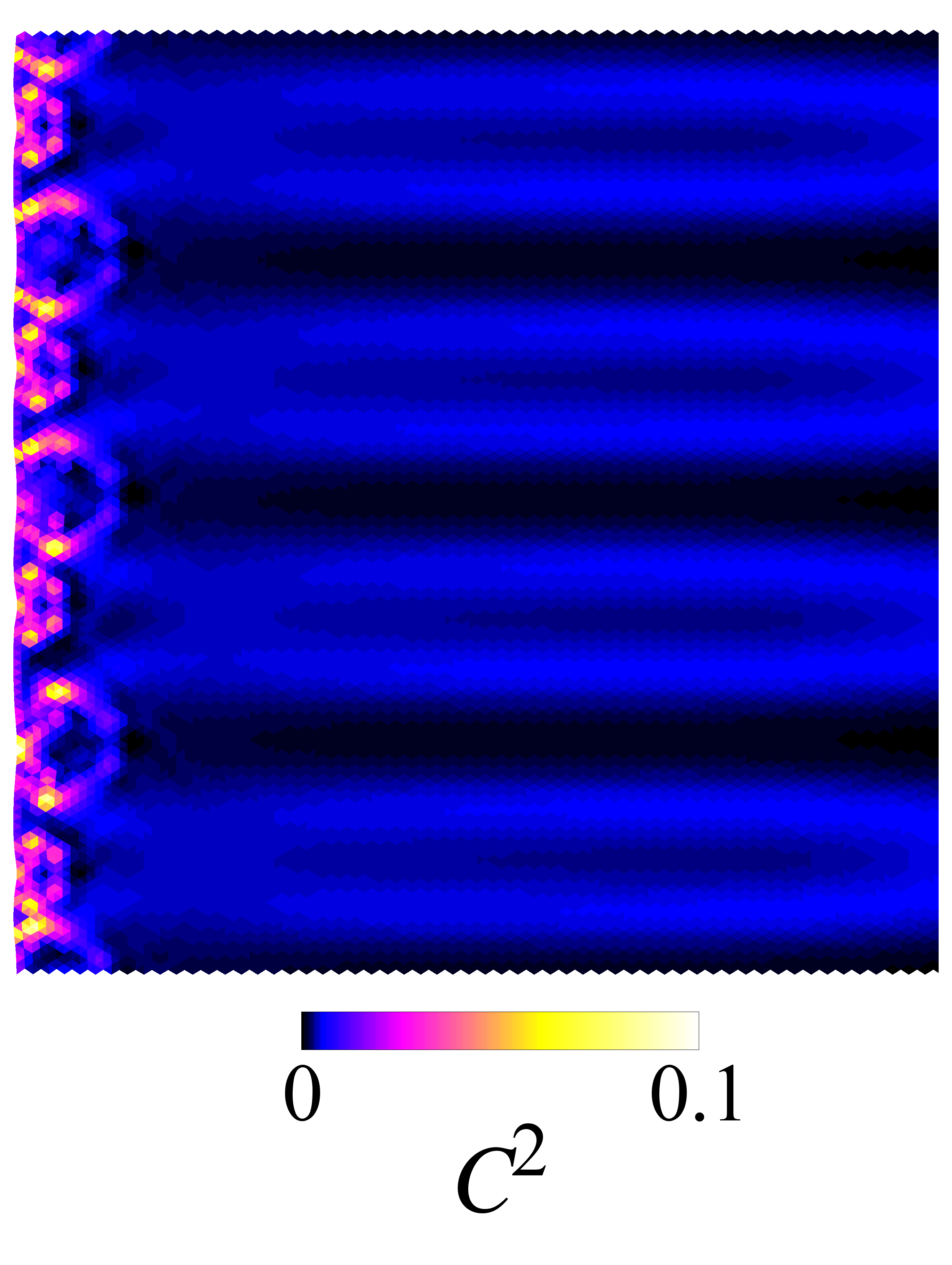} 
	\end{tabular}
	\caption{The solvent fraction and  squared curvature in the isotropic domain of a cylindrical shell with $h_0=0.05$, $R_0=10$ and $L_0=100$. The cylindrical surface is cut along a generatrix for a full view.}
	\label{fig:cyl}
\end{figure}  
   
\begin{figure}[t]
	\begin{tabular}{cc}
		(a) & (b)\\
		\includegraphics[width=0.22\textwidth]{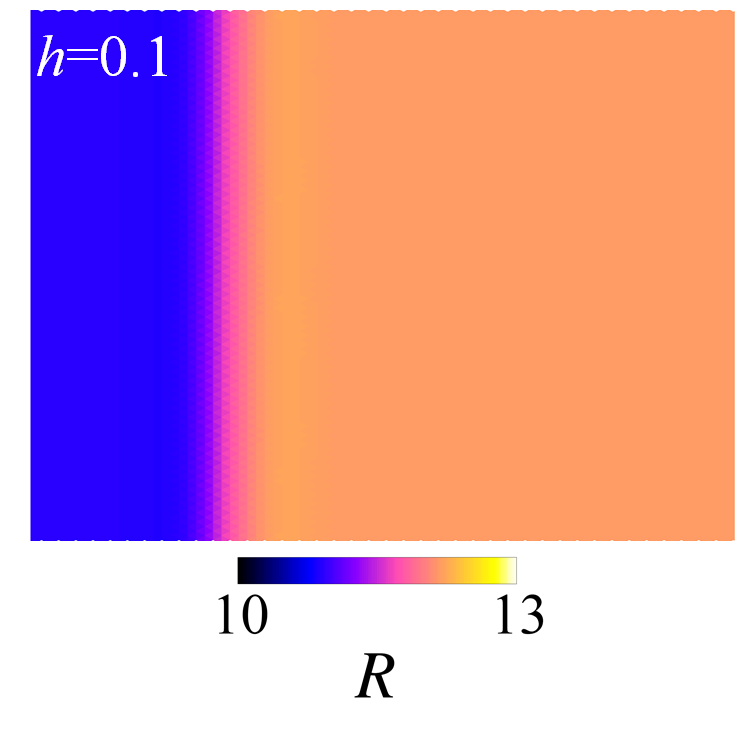}
		&\includegraphics[width=0.22\textwidth]{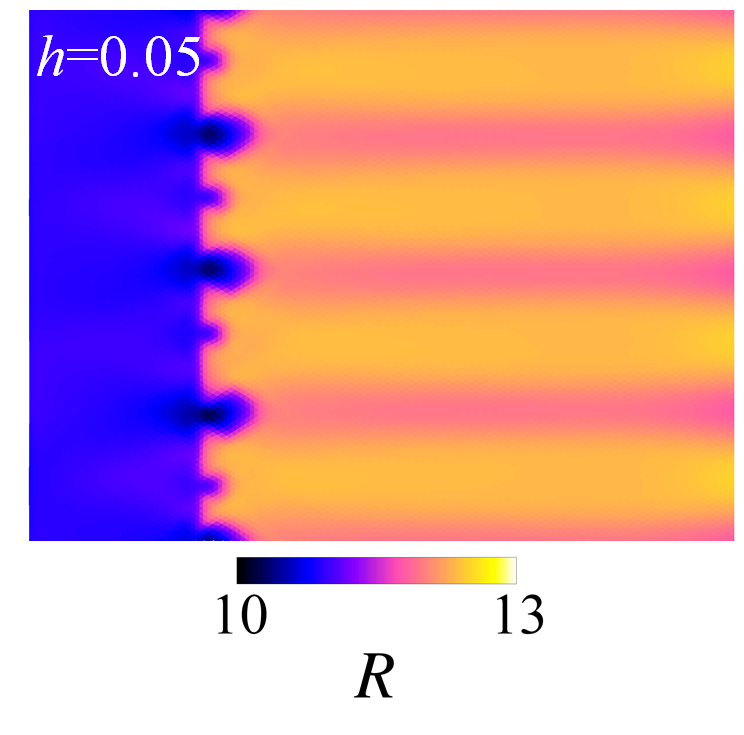}\\
		(c) & (d)\\
		\includegraphics[width=0.22\textwidth]{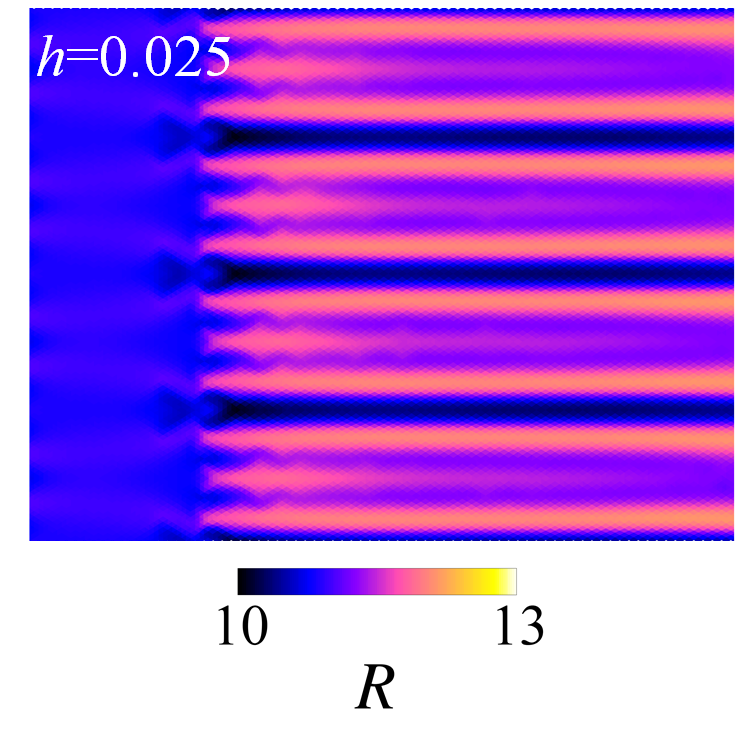}
		&\includegraphics[width=0.22\textwidth]{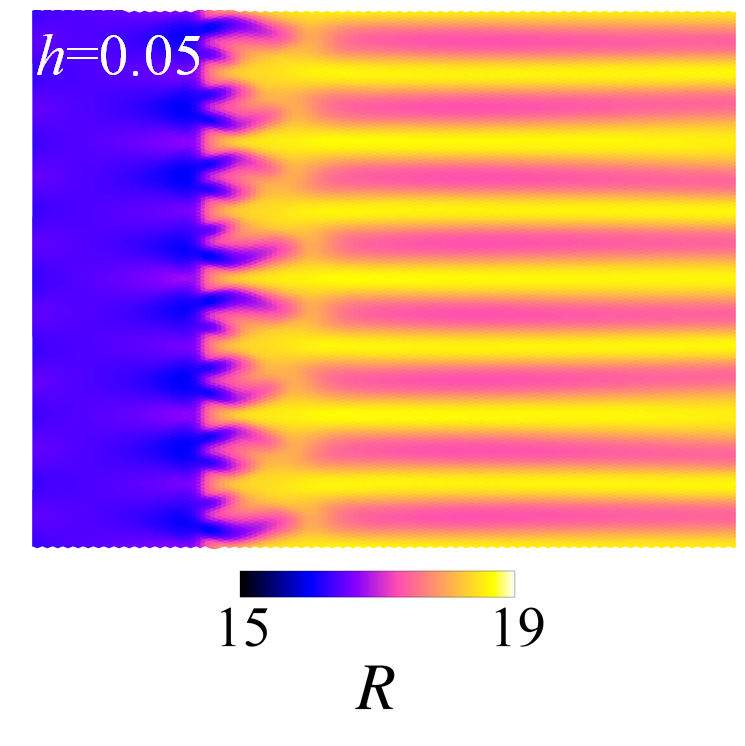}
	\end{tabular}
	\caption{The folding patterns (shown by the changes of the local radius) in cylindrical shells with $L_0=100$,  $R_0=10$ (a,b,c) and $R_0=15$ (d), and different thicknesses (as indicated).  The surface is cut along a generatrix for a full view.}
	\label{fig:cyl_h}
\end{figure}
 
The formation of folds becomes better ordered on a cylindrical shell with a circular front, since the spectrum of admissible wavelengths is limited there by the periodicity around the circumference. A circular front, being a geodesic of the cylindrical surface, has the same structure as a rectilinear front on a flat sheet and, similar to the latter, is neutrally stable to shifts in the normal direction. The correlation between the local curvature and solvent fraction becomes here even more pronounced, as seen in Fig.~\ref{fig:cyl}.

The number of folds increases with decreasing thickness (Fig.~\ref{fig:cyl_h}). In a thin shell one can clearly see that the folds form an hierarchal structure, with the maximum number near the front and decreasing by stages due to convergence of pairs of folds. This process becomes, however, very slow with increasing separation, as it commonly happens in coarsening. The persistence of folds on a cylinder at large distances from the interphase boundary, as compared with their fast decay on a sheet, is a consequence of quantization imposed by the circumferential periodicity. The minimal wavelength of folds is of the same order of magnitude as the front thickness, and the maximum number of folds increases with increasing ratio of the circumference of the shell to the front thickness. 
  
 \subsection{Deformation of a sphere}\label{S33}
 
\begin{figure}[b]
	\begin{tabular}{c}
		(a)  \\
		 \includegraphics[width=0.4\textwidth]{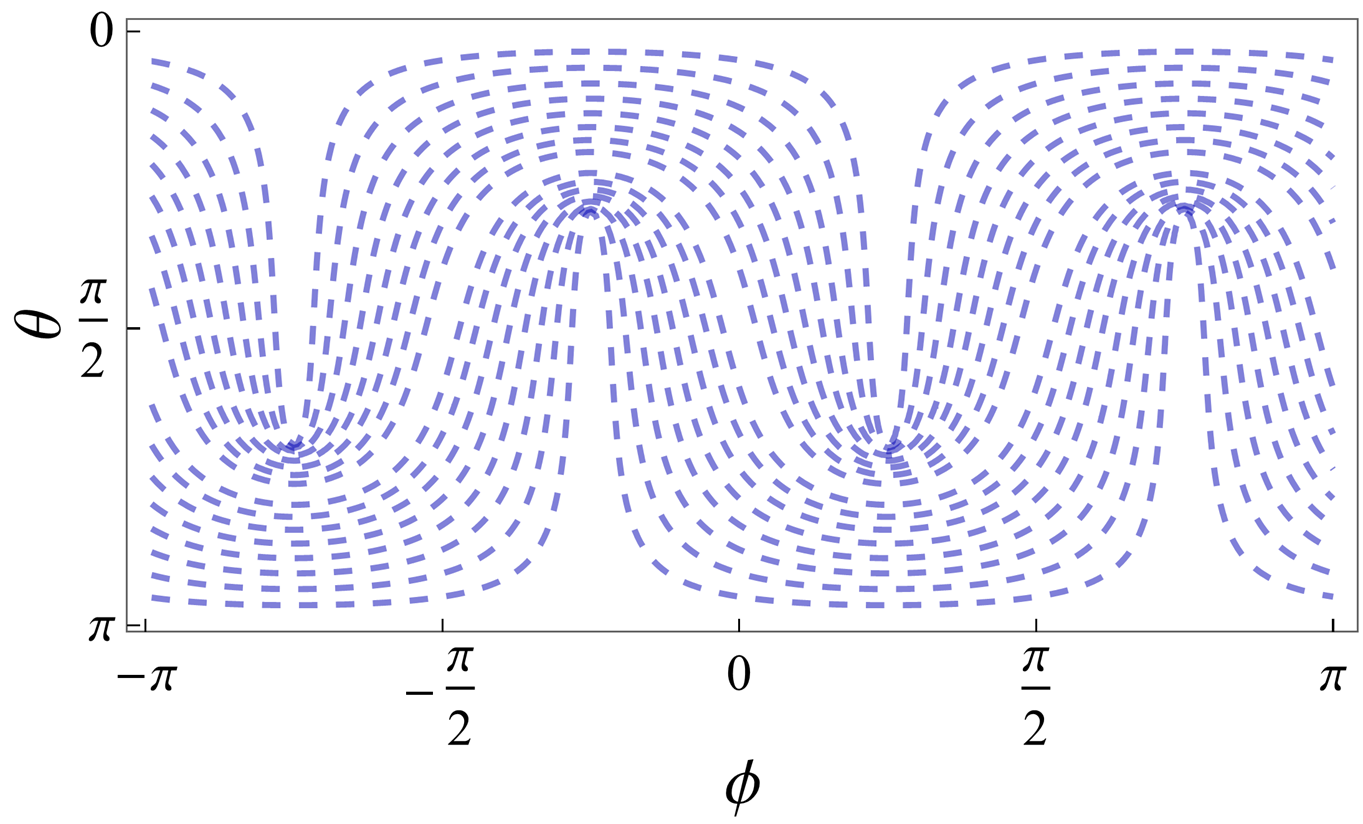} 
		\\ (b)   \\
		\includegraphics[width=0.4\textwidth]{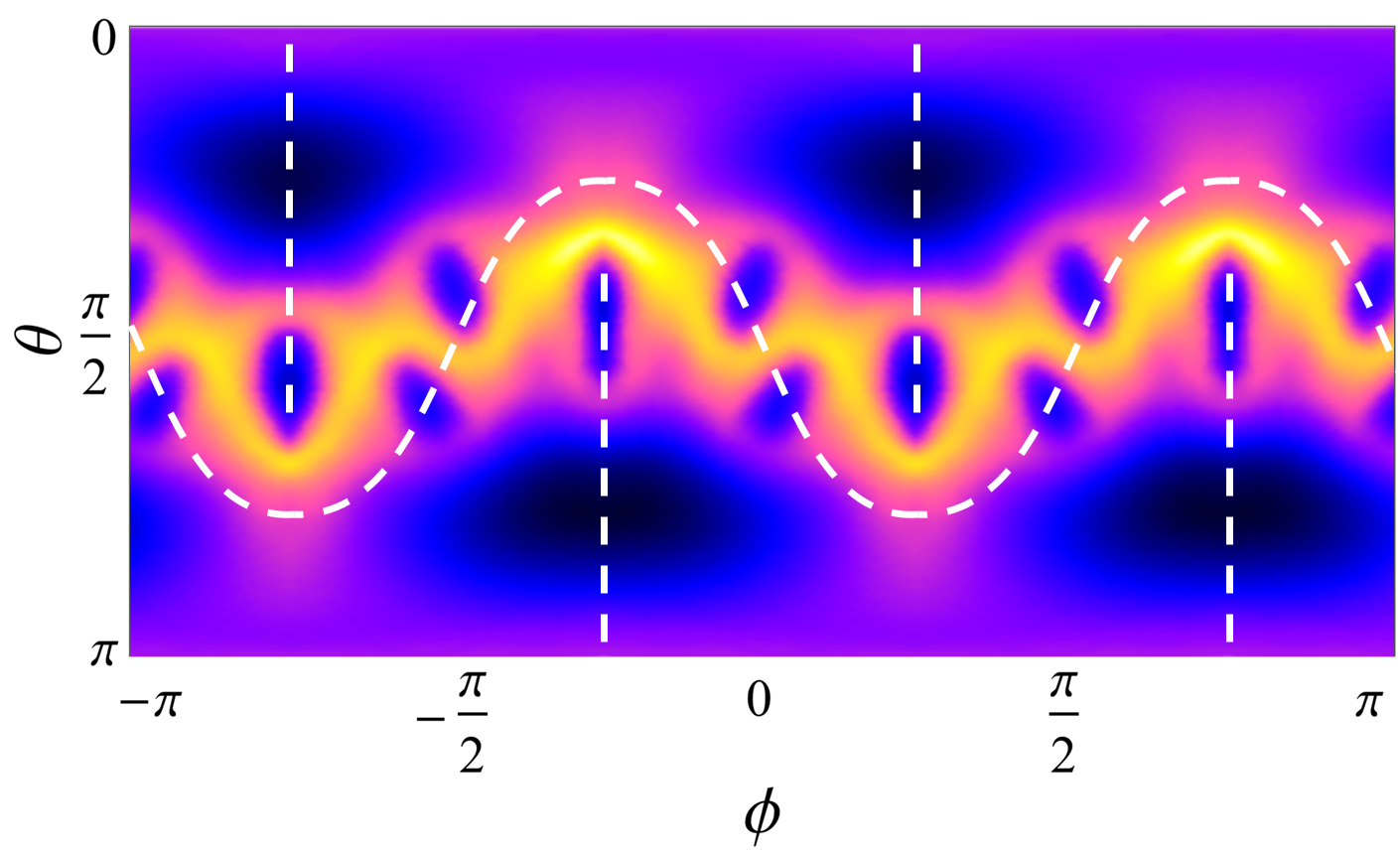} 
\end{tabular}
 	  \begin{tabular}{cc}
  (c) & (d)\\
	  \includegraphics[width=0.24\textwidth]{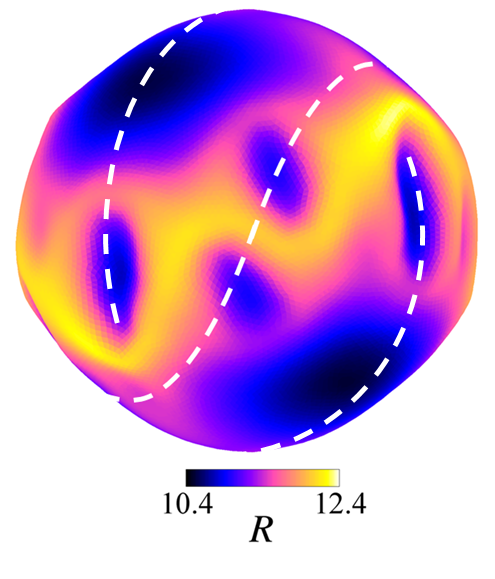}  
	   &\includegraphics[width=0.22\textwidth]{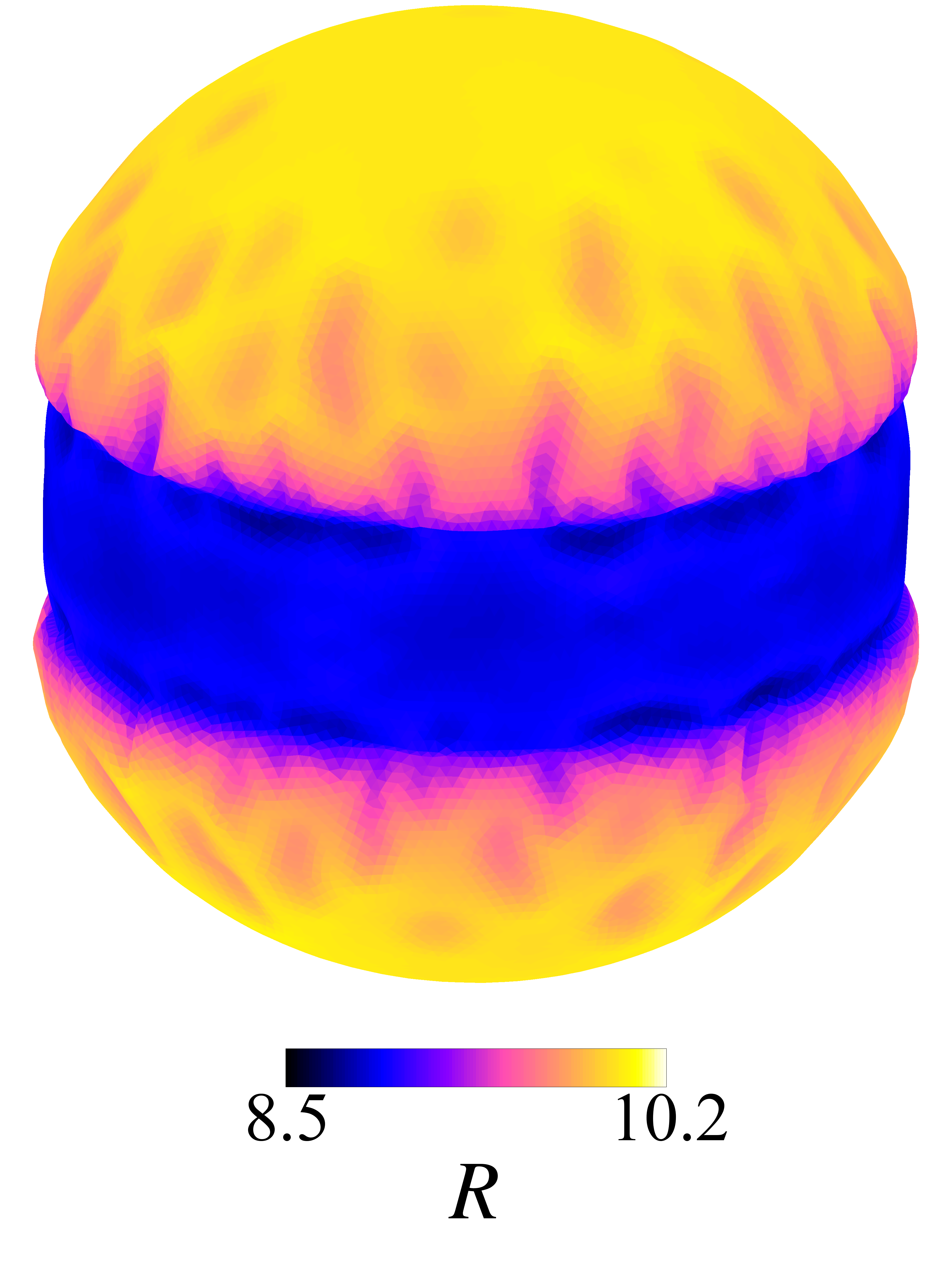} 
	  \end{tabular}
	\caption{(a)  The nematic texture in a spherical shell (in the Mercator projection). (b), (c) The Mercator projection (b) and shape (c) of a spherical shell with the initial radius $R_0=10$ and thickness $h_0=0.05$ after NIT starting from the state with four $+\frac 12$-charged defects. White dashed lines show the ``baseball seam" and the ``comet tails" of two pairs of defects in the original nematic texture. (d) The shape of the shell with the same initial radius and thickness $h_0=0.01$ following isotropic-nematic transition to a biphasic state with an equatorial monodomain nematic belt with fronts located at the polar angle $\Theta=\frac 12 \pi (1 \pm \frac 16)$. Shading in (b)--(d) encodes the local radius. }
	\label{fig:sphere}
\end{figure}

Reshaping of thin spherical shells due to NIT generates much more elaborate forms than a rounded tetrahedral structure of a nematic vesicle governed by the balance of nematic and mechanical elasticity \cite{Park,Selinger}. The origin of either form is in the structure of the nematic texture on the sphere with the hubs on the four $+\frac 12$-charged defects placed symmetrically at vertices of a tetrahedron. This texture, computed analytically via conformal transformation \cite{nelson}, is presented in the Mercator projection in Fig.~\ref{fig:sphere}a. The defects are situated here at the polar angles $\theta_{1,2}=\arccos (1/\sqrt{3})\approx  0.304\, \pi$ and azimuthal angles $\phi_1=-\frac 14 \pi, \, \phi_2=\frac 34 \pi$ in the northern, and at $\theta_{3,4}=\arccos (-1/\sqrt{3})\approx  0.696\, \pi, \; \phi_3=\frac 14 \pi, \, \phi_4=-\frac 34 \pi$ in the southern hemisphere. The nematic director is aligned with level lines of the real part of the appropriate complex analytic function with the zero level corresponding to the ``baseball seam" seen in Fig.~\ref{fig:sphere}a--c as an undulating white belt. The ``comet tails" of two pairs of defects placed in the northern and southern hemispheres connect through the opposite poles.  

When a spherical shell, originally in the nematic state, imbibes the solvent and turns into the isotropic state with a uniform solvent concentration corresponding to the isotropic minimum in Fig.~\ref{fig:2minima}a, a rich folding pattern develops, seen in the Mercator projection in Fig.~\ref{fig:sphere}b and in the side view in Fig.~\ref{fig:sphere}c. Deformation in the vicinity of defects, with folds parallel to their ``comet tails" and a bulge on the other side are similar to that originating from a planar pattern \cite{epj15}. This folding pattern spreads out along the ``baseball seam", while relatively deep depressions appear along the connections between pairs of defects. The deformation structure originates in anisotropic expansion and contraction following NIT that corresponds to the original nematic texture. A plain rounded tetrahedral form with suppressed folding was, however, recovered in our test computation with the shell thickness increased to $h_0=0.75$.
 
Computation of  phase-separated equilibrium states on a sphere is impeded, first, by the absence of a monodomain nematic state on a spherical surface and second, by the absence of geodesics neutrally stable to shifts in the normal direction. The only possible way to set up a monodomain nematic region on a spherical shell is to restrict it to the equatorial belt within the range of polar angles $|\theta-\pi/2| < \Theta$, with the polar regions being isotropic. The isotropic state is likely to nucleate at the defect locations and spread out as the solvent fraction increases, and the final lowest energy state may be eventually achieved, assisted by fluctuations to avoid metastable equilibria. 
 
 Assuming that the front width is small compared to the radius of the sphere and nematic distortions are avoided, the energy of the phase-separated state is expressed as $\mathcal{F}=\mathcal{L}_\mathrm{nem}V_\mathrm{nem}+\mathcal{L}_\mathrm{iso}V_\mathrm{iso}+\sigma L h$, where $\mathcal{L}_\mathrm{nem}, \,\mathcal{L}_\mathrm{iso}$ are energies per unit volume of the uniform nematic and isotropic states, $V_\mathrm{nem}\, V_\mathrm{iso}$ are the volumes they occupy, $\sigma$ is the line tension of the front computed in Sect.~\ref{S21}, and $L, \, h$ are the length of the front and the thickness of the film at this location. The motion of the front is determined by an energy change due to infinitesimal displacements, and therefore its equilibrium is totally determined by local conditions at the front. Since the change of energy due to an infinitesimal displacement $\delta x$ is proportional to the energy difference $\Delta \mathcal{L}= \mathcal{L}_\mathrm{nem}-\mathcal{L}_\mathrm{iso}$ the equilibrium condition is simply $\Delta \mathcal{L} \delta A +\sigma \delta L=0$. On the boundary circle, the equilibrium condition reduces to $\sigma/\Delta \mathcal{L}=\tan \Theta$. This equilibrium is, however, unstable, as the isotropic state would spread further when the front shifts towards the equator or retreats to a latitude closer to the poles. Stable equilibrium can be only attained when the total amount of the solvent is limited. 

\begin{figure}[t]
	\begin{tabular}{c}
		(a)  \\
		 \includegraphics[width=0.4\textwidth]{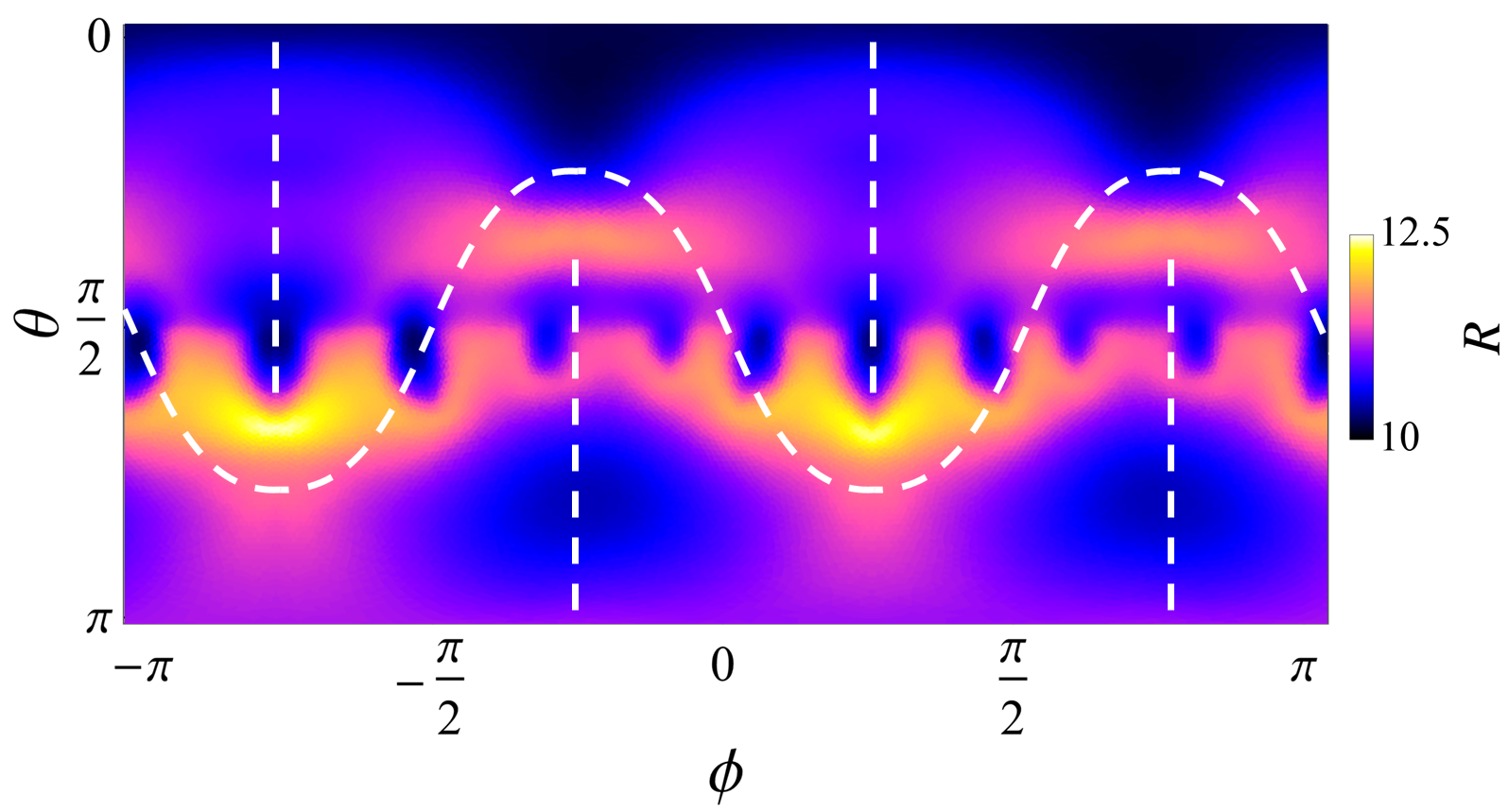} 
		\\ (b)   \\
		\includegraphics[width=0.4\textwidth]{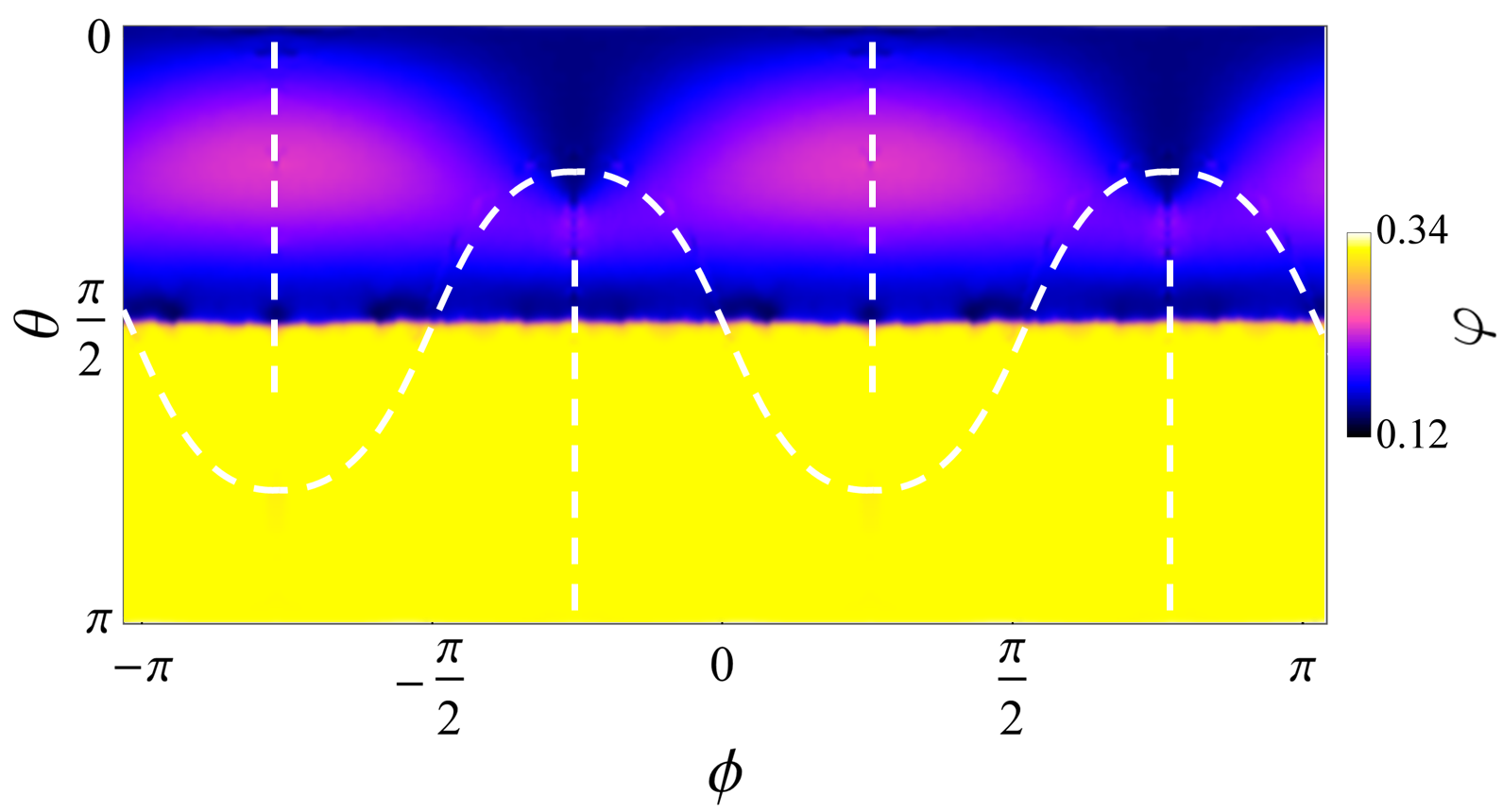} 
\end{tabular}
\caption{The shape (a) and distribution of solvent (b) in a spherical shell with the initial radius $R_0=10$ and thickness $h_0=0.01$ following NIT in the southern hemisphere, with the nematic texture in the northern hemisphere remaining frozen. }
	\label{fig:hemisphere}
\end{figure}

In our commutations, we assume the final configuration to be as described above, with an arbitrary chosen latitude $\Theta$, which fixes the total amount of the solvent.
The equilibrium shape with the uniform meridional alignment in the nematic domain obtained after deforming from the \emph{isotropic} state after partially squeezing out the solvent, shown in Fig.~\ref{fig:sphere}d, is rather bland. There is an expected constriction in the equatorial belt caused both by the loss of solvent and shortening by the factor $\sqrt{1+aS}$. Elongation of the equatorial belt is counteracted by the loss of solvent. Folds develop in the isotropic region and, as usual, are most pronounced near the front. Further toward the poles, they undulate in the meridional direction, unlike straight folds on a cylindrical shell. 

Shapes originating from the nematic state may also evolve to a monodomain equatorial belt after long evolution but the emerging deformation pattern will strongly depend on the orientation of this belt relative to the original pattern and inclusion of original defects. If, on the other hand, the texture in the nematic domain remains frozen, the deformation pattern in the emerging isotropic domain (placed in the southern hemisphere in Fig.~\ref{fig:hemisphere}) is similar to that in the case when NIT takes place in the entire shell, while deformation in the domain remaining in the nematic state is minimal, and is largely caused by a slight increase of solvent content, which takes place on the ``comet tails" of extinct defects extending into the northern hemisphere.    


Different shapes would be obtained starting from a nematic shell with half-charged defects replaced by defects of unit charge, which is possible at certain ratios between splay and bend nematic elasticities \cite{Bowick08prl} or fixing defects at specific locations during polymerization \cite{Poon}.

 \section{Conclusion}

The above shapes present just a small sample of a variety of shapes that can be obtained in nematic elastomers with variable distribution of admixtures (solvent or dopants) affecting the nematic order. The folding patterns emerging due to differential extension or contraction can be compared with folding and wrinkling patterns of different physical origin in soft materials \cite{Sharon,Cerda,Gao,Menon} but their distinguished feature is, on the one hand, anisotropy specific to soft nematic solids and, on the other hand, spatial inhomogeneity that allows one to manipulate them by external inputs.  

\emph{Acknowledgement} This research is supported by Israel Science Foundation  (grant 669/14).



\begin{thebibliography}{99}

\bibitem{degennes75}  P. G. de Gennes, C. R. Seances Acad. Sci. Ser. B \textbf{281}, 101 (1975).

\bibitem{Finkelmann81} H. Finkelmann, H. Kock and G. Rehage, Macromol. Rapid Commun. \textbf{2}, 317 (1981).

\bibitem{Warner88} M. Warner, K. P. Gelling, and T. A. Vilgis, J. Chem. Phys. 88,
4008 (1988).

\bibitem{Khokhlov} S. S. Abramchuk and A. R. Khokhlov, Dokl. Akad. Nauk SSSR
\textbf{297}, 385 (1987).

\bibitem{Pleiner} P. Martinoty, P. Stein, H. Finkelmann, H. Pleiner, and H. R. Brand, Eur. Phys. J. E  \textbf{14}, 311 (2004).

\bibitem{Warner} M. Warner and E. M. Terentjev, \emph{Liquid Crystal Elastomers} (Clarendon Press, Oxford, 2003).

\bibitem{Warner12} C. D. Modes, K. Bhattacharya, and M. Warner, Proc. R. Soc. A  \textbf{467}, 1121 (2011).

\bibitem{sharon} H. Aharoni, E. Sharon, and R. Kupferman, Phys. Rev. Letts. \textbf{113}, 257801 (2014).

\bibitem{Cirak14} F. Cirak, Q. Long, K. Bhattacharya, and M. Warner, Inte. J. Solids and Structures \textbf{51}, 144 (2014).

\bibitem{epj15} A. P. Zakharov and L. M. Pismen, Eur. Phys. J. E  \textbf{38}, 75 (2015).

\bibitem{mostajeran2015curvature}C. Mostajeran, Phys. Rev. E \textbf{91}, 062405 (2015).

\bibitem{mostajeran2016encoding} C. Mostajeran, M. Warner, T. H. Ware, and T.J . White, Proc. R. Soc. A \textbf{472}, 20160112 (2016).

\bibitem{Broer14} L. T. de Haan, A.P.H.J. Schenning, and D. J. Broer,  Polymer \textbf{55}, 5885 (2014).

\bibitem{BroerLa} D. Liu and D. J. Broer, Langmuir, \textbf{30}, 13499 (2014).

\bibitem{White} T. H. Ware, M. E. McConney, J. J. Wie, V. P. Tondiglia, and T. J. White, Science \textbf{347}, 982 (2015).

\bibitem{WhiteSM} K. Fuchi, T. H. Ware, P. R. Buskohl, G. W. Reich, R. A. Vaia, T. J. White. and J. J. Joo, Soft Matter \textbf{11}, 7288 (2015).

\bibitem{Finkelmann91} J. Kupfer and H. Finkelmann, Makromol. Chem. Rapid Comm.
12, 717 (1991).

\bibitem{Ikeda} T. Ikeda,  J. Mater. Chem. \textbf{13}, 2037 (2003).

\bibitem{Samitsu} S. Samitsu, Y. Takanishi, and J. Yamamoto, Nature Materials, \textbf{9}, 816 (2010). 

\bibitem{Broer} D. Liu and D. J. Broer, Nature Comm..\textbf{6}, 8334  (2015).

\bibitem{UrayamaN} K. Urayama, Y. Okuno, T. Kawamura, and S. Kohjiya, Macromolecules, \textbf{35}, 4567  (2002).

\bibitem{Urayama03} K. Urayama, Y. Okuno, and S. Kohjiya, Macromolecules, \textbf{36}, 6229  (2003).

\bibitem{Urayama06} K. Urayama, R. Mashita, Y. O. Arai, and T. Takigawa, Macromolecules, \textbf{39}, 8511  (2006).

\bibitem{Bahar} I. Bahar and B. Erman, Macromolecules, \textbf{20}, 1696  (1987).

\bibitem{Terentjev2015} N. Cheewaruangroj and E. M. Terentjev, Phys. Rev. E, \textbf{92}, 042502 (2015).

\bibitem{Fukuda} J. Fukuda, Phys. Rev. E \textbf{58}, R6939 (1998).

\bibitem{rey04} S. K Das and A. D. Rey, J. Chem. Phys. \textbf{121}, 9733 (2004).

\bibitem{kopf2013phase} M. H. K{\"o}pf and  L.M. Pismen, Eur. Phys. J. E  \textbf{36}, 1 (2013).

\bibitem{arxivorg} A. P. Zakharov and L. M. Pismen, 
\\ https://arxiv.org/abs/1608.03942 (2016).

\bibitem{Flory}P. J. Flory,  J. Chem. Phys. \textbf{10}, 51 (1942).

\bibitem{Delaunay} F. P. Preparata and M. Shamos, \emph{Computational geometry: an introduction} (Springer, 2012).

\bibitem{Park} J. Park, T. C. Lubensky. and F. C. MacKintosh, Europhys. Lett., \textbf{20}, 279 (1992).

\bibitem{Selinger} T.-S. Nguyen , J. Geng  , R. L. B. Selinger, and J. V. Selinger, Soft Matter, \textbf{9}, 8314 (2013).

\bibitem{nelson} V. Vitelli and D. R. Nelson, Phys. Rev. E \textbf{74}, 021711 (2006).

\bibitem{Bowick08prl} H. Shin, M. J. Bowick, and X. Xing, Phys. Rev. Lett. \textbf{101}, 037802 (2008).

\bibitem{Poon} J. H. Noh, B. Henx, and J. P. F. Lagerwall, Adv. Mater. \textbf{28}, 10170 (2016).

\bibitem{Sharon} Y. Klein, E. Efrati and E. Sharon, Science, \textbf{315}, 1116  (2007).

\bibitem{Cerda} L. Pocivavsek, R. Dellsy, A. Kern, S. Johnson, B. Lin, K. Yee, C. Lee, and E. Cerda, Science \textbf{320},  912 (2008).

\bibitem{Gao} B. Li, Y.-P. Cao,  X.-Q. Feng, and   H. Gao, Soft Matter \textbf{8}, 5728 (2012) 

\bibitem{Menon} J. Huang, B. Davidovitch, T. P. Russell, and N. Menon,
Phys. Rev. Lett. \textbf{105},  038302 (2010) 

\end{thebibliography}
\end{document}